\documentclass[12pt,preprint]{aastex}

\newcommand\etal{{et al. }}
\def\farcs{\hbox{$.\!\!^{\prime\prime}$}}

\slugcomment{To be published in The Astrophysical Journal}

\shorttitle{Dust and SN~2003gd}
\shortauthors{Meikle et al.}

\begin{document}
%% LaTeX will automatically break titles if they run longer than
%% one line. However, you may use \\ to force a line break if
%% you desire.
\title{A Spitzer Space Telescope Study of SN~2003gd: Still No Direct
Evidence that Core-Collapse Supernovae are Major Dust Factories}
%% Use \author, \affil, and the \and command to format
%% author and affiliation information.
%% Note that \email has replaced the old \authoremail command
%% from AASTeX v4.0. You can use \email to mark an email address
%% anywhere in the paper, not just in the front matter.
%% As in the title, use \\ to force line breaks.

\author{W. P. S. Meikle,\altaffilmark{1}
S. Mattila,\altaffilmark{2} 
A. Pastorello,\altaffilmark{2} 
C. L. Gerardy,\altaffilmark{1} 
R. Kotak,\altaffilmark{2}
J. Sollerman,\altaffilmark{3} 
S.~D.~Van~Dyk,\altaffilmark{4} 
D.~Farrah,\altaffilmark{5} 
A. V. Filippenko,\altaffilmark{6} 
P. H\"oflich,\altaffilmark{7},
P. Lundqvist,\altaffilmark{8} 
M. Pozzo,\altaffilmark{9}  and
J.~C.~Wheeler\altaffilmark{10}} 

\altaffiltext{1}{Astrophysics Group, Blackett Laboratory, Imperial
College London, Prince Consort Road, London SW7 2AZ, United Kingdom;
p.meikle@imperial.ac.uk, c.gerardy@imperial.ac.uk}
\altaffiltext{2}{Astrophysics Research Centre, School of Mathematics and
Physics, Queen's University Belfast, BT7 1NN, United Kingdom;
s.mattila@qub.ac.uk, a.pastorello@qub.ac.uk, r.kotak@qub.ac.uk}
\altaffiltext{3}{Dark Cosmology Centre, Niels Bohr Institute,
University of Copenhagen, Juliane Maries Vej 30, 2100 Copenhagen,
Denmark; jesper@astro.su.se}
\altaffiltext{4}{Spitzer Science Center/Caltech, Mail Code 220-6,
Pasadena, CA 91125; vandyk@ipac.caltech.edu}
\altaffiltext{5}{Department of Astronomy, 106 Space Sciences Building,
Cornell University, Ithaca, NY 14853; duncan@isc.astro.cornell.edu}
\altaffiltext{6}{Department of Astronomy, University of
California, Berkeley, CA 94720-3411; alex@astro.berkeley.edu}
\altaffiltext{7}{Department of Physics, Florida State University,
Tallahassee, FL 32306; pah@astro.physics.fsu.edu}
\altaffiltext{8}{Stockholm University, AlbaNova University Center,
Stockholm Observatory, Department of Astronomy, SE-106 91 Stockholm,
Sweden; peter@astro.su.se}
\altaffiltext{9}{Department of Earth Sciences, University College
London, Gower Street, London WC1E 6BT, UK; m.pozzo@ucl.ac.uk}
\altaffiltext{10}{The University of Texas at Austin, Department of
Astronomy, Austin, TX 78712; wheel@astro.as.utexas.edu}

%% Notice that each of these authors has alternate affiliations, which
%% are identified by the \altaffilmark after each name.  Specify alternate
%% affiliation information with \altaffiltext, with one command per each
%% affiliation.

%\altaffiltext{1}{Visiting Astronomer, Cerro Tololo Inter-American Observatory.
%CTIO is operated by AURA, Inc.\ under contract to the National Science
%Foundation.}
%\altaffiltext{2}{Society of Fellows, Harvard University.}
%\altaffiltext{3}{present address: Center for Astrophysics,
%    60 Garden Street, Cambridge, MA 02138}
%\altaffiltext{4}{Visiting Programmer, Space Telescope Science Institute}
%\altaffiltext{5}{Patron, Alonso's Bar and Grill}

%% Mark off your abstract in the ``abstract'' environment. In the manuscript
%% style, abstract will output a Received/Accepted line after the
%% title and affiliation information. No date will appear since the author
%% does not have this information. The dates will be filled in by the
%% editorial office after submission.

\begin{abstract}
We present a new, detailed analysis of late-time mid-infrared (IR)
observations of the Type~II-P supernova (SN) 2003gd. At about 16~months
after the explosion, the mid-IR flux is consistent with emission from
$4\times10^{-5}$ M$_{\odot}$ of newly condensed dust in the ejecta.
At 22~months emission from point-like sources close to the SN position
was detected at 8~$\mu$m and 24~$\mu$m. By 42~months the 24~$\mu$m
flux had faded.  Considerations of luminosity and source size rule out
the ejecta of SN~2003gd as the main origin of the emission at
22~months.  A possible alternative explanation for the emission at
this later epoch is an IR~echo from pre-existing circumstellar or
interstellar dust.  We conclude that, contrary to the claim of
\citet{sug06}, the mid-IR emission from SN~2003gd does not support the
presence of 0.02~M$_{\odot}$ of newly formed dust in the ejecta.
There is, as yet, no direct evidence that core-collapse supernovae are
major dust factories.
\end{abstract}

%% Keywords should appear after the \end{abstract} command. The uncommented
%% example has been keyed in ApJ style. See the instructions to authors
%% for the journal to which you are submitting your paper to determine
%% what keyword punctuation is appropriate.

\keywords{supernovae: general ---
supernovae: individual (\objectname{SN 2003gd})}

%% From the front matter, we move on to the body of the paper.
%% In the first two sections, notice the use of the natbib \citep
%% and \citet commands to identify citations.  The citations are
%% tied to the reference list via symbolic KEYs. The KEY corresponds
%% to the KEY in the \bibitem in the reference list below. We have
%% chosen the first three characters of the first author's name plus
%% the last two numeral of the year of publication as our KEY for
%% each reference.

%% Authors who wish to have the most important objects in their paper
%% linked in the electronic edition to a data center may do so by tagging
%% their objects with \objectname{} or \object{}.  Each macro takes the
%% object name as its required argument. The optional, square-bracket 
%% argument should be used in cases where the data center identification
%% differs from what is to be printed in the paper.  The text appearing 
%% in curly braces is what will appear in print in the published paper. 
%% If the object name is recognized by the data centers, it will be linked
%% in the electronic edition to the object data available at the data centers  
%%
%% Note that for sources with brackets in their names, e.g. [WEG2004] 14h-090,
%% the brackets must be escaped with backslashes when used in the first
%% square-bracket argument, for instance, \object[\[WEG2004\] 14h-090]{90}).
%%  Otherwise, LaTeX will issue an error. 

\section{Introduction}
Massive stars explode via core collapse and ejection of their
surrounding layers (e.g., Arnett et al. 1989, and references therein).
The extent to which core-collapse supernovae (CCSNe) are, or have
been, a major source of dust in the universe is of great interest.
For many years it has been hypothesized that the physical conditions
in the expanding ejecta of CCSNe could result in the condensation of
large masses of dust grains
\citep{cer67,hoy70,geh89,tie90,dwe98,tod01,noz03}.  This follows from
the fact that large abundances of suitable refractory elements are
present. In addition, cooling by adiabatic expansion and molecular
emission takes place, and dynamical instabilities can produce density
enhancements or ``clumping.''  This, in turn, will aid dust formation
through the effects of cooling and self-shielding.  Further support
for these ideas is provided by isotopic anomalies in meteorites which
indicate that some grains must have formed in CCSNe \citep{cla97}. \\

Interest in CCSNe as dust producers has increased recently due to the
problem of accounting for the presence of dust at high redshifts
\citep{fal89,fal96,pei91,pet97,ber03}.  In these early eras, much less
dust production from novae and asymptotic giant branch stars is
expected since fewer stars will have evolved past the main-sequence
phase.  Consequently, CCSNe arising from Population~III stars have
been proposed as the main early-universe source of dust
\citep{tod01,noz03}.  Models of dust formation in CCSNe
\citep{tod01,noz03} succeed in producing copious amounts of dust ---
around 0.1--1~M$_{\odot}$ even in the low-metallicity environments at
high redshifts.  This corresponds to a supernova (SN) dust
condensation efficiency of about 0.2 \citep{mor03}, where the
efficiency is defined as the dust mass divided by the total mass of
refractory elements.  This is enough to account for the quantity of
dust seen at high redshifts (see Appendix). \\

Newly condensed dust in CCSNe can be detected by its attenuating
effects on optical/near-infrared (IR) light or via thermal emission
from the grains in the ejecta.  These methods have been used in
attempts to measure the dust productivity of CCSNe.  Both methods are
subject to uncertainties due to dust formation in optically thick
clumps, so the derived masses tend to be just lower limits.  By far
the most extensive evidence for ejecta dust condensation is that
obtained from the peculiar Type~II SN~1987A, where both techniques
were employed \citep{dan89,luc89,mei89,whi89,sun90,dwe92,roc93,woo93,erc07}.
However, even the highest value obtained is only
$7.5\times10^{-4}$~M$_{\odot}$ \citep{erc07}.  \citet{poz04} used the
attenuation method to infer a dust mass exceeding
$2\times10^{-3}$~M$_{\odot}$ in the Type~IIn SN~1998S. However, such
events are relatively rare. Moreover, in this case it is suggested
that the dust condensation was not in the body of the ejecta, but
rather took place in the cool, dense shell produced by the impact of
the SN ejecta with circumstellar material (CSM).  We note also that an
alternative IR-echo scenario for SN~1998S is not ruled out
\citep{ger02,poz04}.  \\

Prior to the launch of the {\it Spitzer Space Telescope} (hereafter,
{\it Spitzer}), the only evidence of dust condensation in a typical
CCSN was presented by \citet{elm03}, who used optical line attenuation
to infer a dust mass lower limit of about $10^{-4}$~M$_{\odot}$ in the
Type II-plateau (II-P) SN~1999em.  Mid-IR studies of the Cassiopeia~A
supernova remnant (SNR) \citep{dwe87,lag96,dou01} indicate that dust
formation took place during its explosion, but again the mass of
directly observed dust is small.  Sub-millimeter studies of this SNR
by \citet{dun03} using SCUBA led them to claim that at least
2~M$_\odot$ of dust formed in the supernova.  However, \citet{kra04}
have used the same data together with observations from {\it Spitzer}
to show that most of this emission originates from a line-of-sight
molecular cloud, and not from dust formed in Cas~A. \citet{tem06} used
{\it Spitzer} observations to estimate $10^{-3}-10^{-2}$~M$_\odot$ of
dust in the Crab Nebula SNR. While rather uncertain, this result may
be more relevant to this paper than that of Cas~A, since the Crab
Nebula is thought to have arisen from a progenitor of mass
8--10~M$_\odot$ \citep{nom82,kit06}, similar to that of the CCSN
studied here (SN~2003gd). \\

In summary, prior to the launch of {\it Spitzer}, direct observations
of CCSNe or SNRs have never revealed more than
$\sim$10$^{-3}$~M$_{\odot}$ of dust --- only $\sim$1\% of the mass
required if CCSNe are to be important dust sources.  But the number of
CCSNe investigated for dust production is small, and with the
exception of SN~1999em, rather atypical.  The availability of {\it
Spitzer} has provided an excellent opportunity for us to test the
ubiquity of dust condensation in a statistically significant number of
typical CCSNe.  It provides high-sensitivity imaging over the mid-IR,
covering the likely peak of the dust thermal emission spectrum. This
can provide a superior measure of the total flux, temperature and,
possibly, dust emissivity than can be achieved at shorter wavelengths.
Moreover, the longer-wavelength coverage of {\it Spitzer} lets us
detect cooler grains and see more deeply into dust clumps than was
previously possible for typical nearby CCSNe.  In addition,
multi-epoch observations with {\it Spitzer} can distinguish between
dust condensation and IR~echoes via the strength and shape of the
light curve. \\

In this paper we analyze {\it Spitzer} observations of the Type~II-P
SN~2003gd at three late-time epochs. \citet{hen05} and \citet{sug06}
(henceforth ``S06'') reported optical attenuation effects in the
late-time spectra and $BR$ light curves of SN~2003gd which indicate
dust condensation in this event.  Using {\it Spitzer} observations at
two late-time epochs, S06 also report mid-IR emission from the
condensing dust.  This was the first-ever report of condensing dust in
a SN~II-P on the basis of thermal emission from the grains. Here we
present a new study of these {\it Spitzer} observations.  We agree
with S06 that some of the earlier-epoch mid-IR emission was due to a
modest quantity of ejecta dust.  However, we find that their principal
conclusion, that the later-epoch observations indicate the presence of
0.02~M$_\odot$ of dust formed in the ejecta, is not supported by the
data.  Consequently, thus far there is no direct evidence that CCSNe
are major dust factories.

\section{Observations}
SN~2003gd was discovered \citep{eva03} on 2003 June 12 (UT dates are
used throughout this paper) in the SA(s)c galaxy NGC~628 (M74).  On
2003 June 13 it was identified as a Type~II event \citep{gar03} using
a $J$-band spectrum. On 2003 June 14 the identification was confirmed
using optical spectra, and it was estimated that the SN was roughly
1~month \citep{phi03} or 2~months \citep{kot03} post-explosion at the
time of discovery.  Using light-curve comparison with other SNe~II-P,
it was deduced \citep{vdy03,hen05} that SN~2003gd was a normal
Type~II-P event with estimated explosion dates of, respectively, 2003
March $17\pm3$ \citep{vdy03} or $18\pm21$ \citep{hen05}.  We adopt
2003 March~17 as the explosion date, 87~days pre-discovery.  \\

On the basis of a variety of methods (standardized candle method,
brightest supergiants, kinematic) a distance to SN~2003gd of
$9.3\pm1.8$~Mpc was found \citep{hen05}. Modelling of the light echo
of SN~2003gd \citep{vdy06} suggests a somewhat smaller distance of
about 7~Mpc. S06 adopted 9.3~Mpc, and so for ease of comparison with
their work we shall adopt 9.3~Mpc throughout.  Total extinction
(Galactic + host galaxy) estimates of $E(B-V)=0.13 \pm 0.03$~mag
\citep{vdy03} and $E(B-V)=0.14\pm0.06$~mag \citep{hen05} were
reported.  Using two independent methods (bolometric luminosity of
exponential tail; direct comparison with SN~1987A bolometric light
curve), Hendry \etal estimate an ejected $^{56}$Ni mass of
0.016$^{+0.010}_{-0.006}$~M$_{\odot}$, only about a fifth of the
$^{56}$Ni mass found in SN~1987A \citep{whi88,bou91}.  The progenitor
star was identified in archival images from the {\it Hubble Space
Telescope}, the 2.6-m Nordic Optical Telescope, and Gemini North, as a
red supergiant of mass 6--12~M$_{\odot}$ \citep{vdy03,sma04}.\\

The field of SN~2003gd was observed with {\it Spitzer's} Infrared
Array Camera ({\it IRAC}) at 3.6, 4.5, 5.8, and 8.0~$\mu$m on 2004
July 25 and 28 (days 496 and 499) and again on 2005 January 15
(day~670).  The first two SN observations were obtained
serendipitously within the {\it Spitzer Infrared Nearby Galaxies
Survey (SINGS)} (PID: 0159) \citep{ken03}.  In the {\it SINGS} program
observations are duplicated with a delay of a few days, to permit
identification of asteroids and to better sample the emission on
subpixel scales \citep{reg04}.  In each wavelength channel the two
images are combined to yield an ``Enhanced Data Product,'' and these
are publicly available from the NED database.  The 2005 January
observation was obtained within our {\it Spitzer} supernova program
(PID: 3248).  The {\it SINGS} program also used the {\it Multiband
Imaging Spectrometer for Spitzer (MIPS)} to acquire images of the
field of SN~2003gd at 24~$\mu$m on 2005 January 23 and 26 (days 678
and 681). In our measurement and analysis of the days 496/9, 670, and
678/81 observations, we used the same data as were available to S06.
SN~2003gd was again observed at 24~$\mu$m within the {\it Spitzer}
supernova program of Sugerman \etal (PID:~30494) on 2006~September~1
(day~1264). Subsequent to the initial submission of this paper, Dr. Ben
Sugerman kindly made this image available to us. We therefore also
consider the implications of this observation.

\section{Results}
\begin{figure*}
\epsscale{0.87}
\plotone{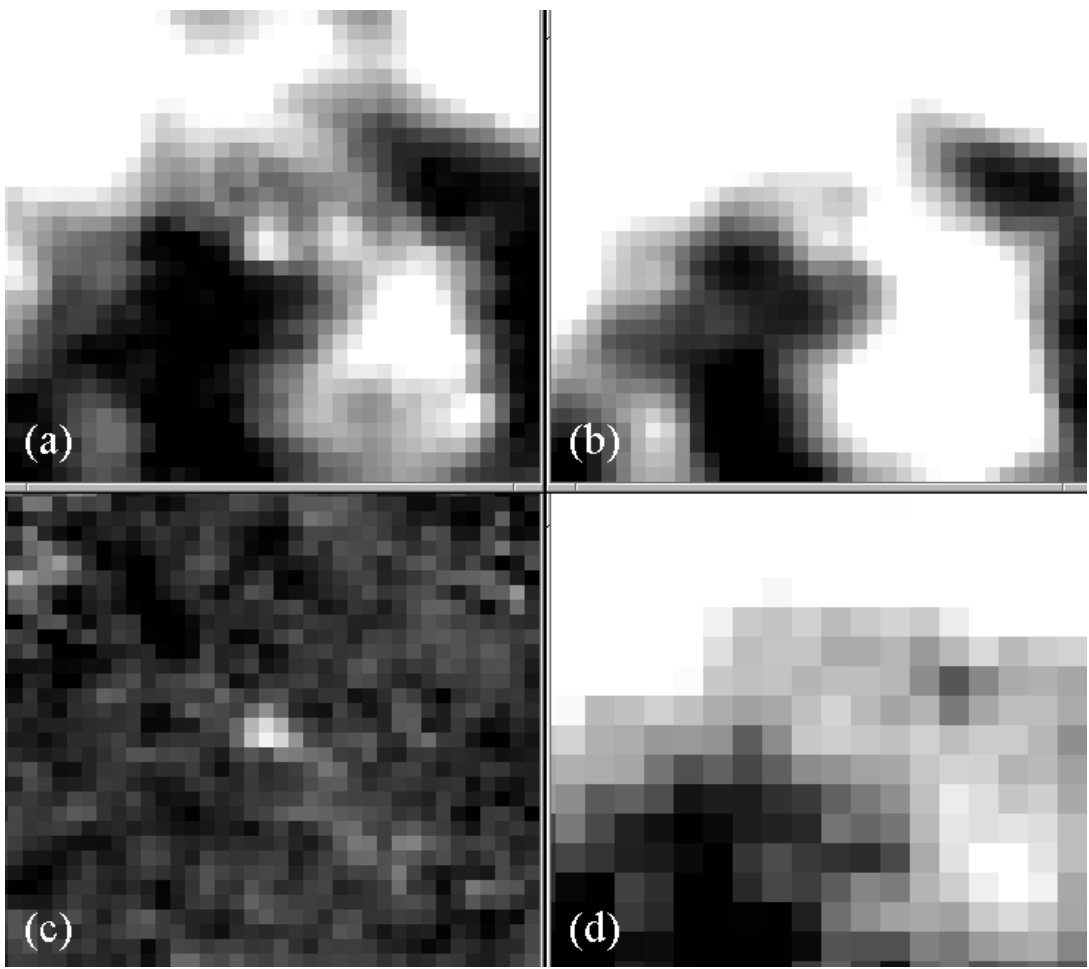}
\caption[]{Panels (a) and (b) show the field of SN~2003gd at 8~$\mu$m
({\it IRAC}) on day~496/9 and day~670, respectively.  Panel (c) shows
the result of subtracting (b) from (a). The bright point source is
SN~2003gd.  Panel (d) shows the field of SN~2003gd at 24~$\mu$m ({\it
MIPS}) on day~678/81.  North is up, and east is to the left. The pixel
scales are $0\farcs75$ pixel$^{-1}$ and $1\farcs5$ pixel$^{-1}$ at
8~$\mu$m and 24~$\mu$m, respectively.  Each panel is centred on the
supernova position, determined using PSF fitting and the IRAF GEOMAP
package (see text).
\label{fig1}
}
\end{figure*}

A point source at the SN position is clearly visible in the day~496/9
image in all four {\it IRAC} channels, with strong fading by
day~670. This is illustrated in Figure \ref{fig1}(a,b), where we show
the 8~$\mu$m {\it IRAC} images from day~496/9 and day~670.  Given the
large decline in flux, we deduce that most of the point-source flux
detected on day~496/9 was due to the SN.  However, measurement of the
SN flux is challenging owing to the bright, complex field within which
it lies.

\subsection{Day~496/9 Results}

\subsubsection{PSF-Fitting Measurements}
We used the {\it SNOOPY} point-spread function (PSF) fitting package
to determine the SN fluxes and coordinates.  {\it SNOOPY} was
originally designed by F. Patat to carry out SN photometry. It was
implemented in IRAF by E. Cappellaro and is based on DAOPHOT, and has
been tested and improved over a number of years.  Several suitable PSF
stars are selected in order to build the model PSF and measure the
full-width at half-maximum intensity (FWHM). First a polynomial
surface, of orders between 3 and 6 in $x$ and $y$, is fitted to the
background in a $(10 \times {\rm FWHM}) \times (10 \times {\rm FWHM})$ 
region centered on the SN position, excluding the innermost square region
around the SN, of side $\sim1.5 \times {\rm FWHM}$.  This is then subtracted
from the image. Next the PSF fitting is performed on the SN.  The
fitted PSF is subsequently subtracted from the data to produce a
residual image.  This is inspected by eye, and the fitting procedure
repeated until a residual image is obtained where there is little sign
of the original point source. The code returns the $x$ and $y$
position and the flux within the PSF.  It also provides a statistical
uncertainty which is a measure of how well the model PSF describes the flux
value and distribution at the SN position.  However, the flux values
are quite sensitive to the fitting of the image background with the
polynomial surface.  This may introduce an additional uncertainty in
the absolute flux values, although the effect on the shape of the
spectral energy distribution is likely to be less than this. \\

The results and estimated uncertainties are shown in Table \ref{tab1}.
As a check of the PSF-fitting procedure, field stars were also
measured using both this method and aperture photometry. The aperture
radius was $10\arcsec$ with a 15--20$\arcsec$ concentric sky annulus.
No significant systematic flux difference was found between the two
methods at any wavelength.  The root-mean-square (rms) scatter in the
differences was 0.05--0.18~mag. The rms scatter at each wavelength was
adopted as the uncertainty. As a further check, we performed aperture
photometry for three stars in both post-basic calibrated data (PBCD)
and SINGS-processed {\it IRAC} 8~$\mu$m frames, and found that the
photometry agrees to within 5\%. This test was also applied to the
day~678/81 {\it MIPS} 24~$\mu$m frames (see below) and similar
consistency was obtained.

\subsubsection{Image-Subtraction Measurements}
We also determined the {\it difference} in the fluxes between
day~496/9 and day~670 via image subtraction.  While this only gives
the change in flux between the two epochs, it is a particularly
powerful method of removing the effects of a spatially varying
background such as is encountered in SN images from {\it Spitzer}
\citep{mei06}. Also, given the very weak flux at the SN location on
day~670, this procedure provides a robust check on the net supernova
emission. \\

For each channel, the day~670 image (PBCD processed) was subtracted
from the earlier Enhanced Data Product SINGS image through the use of
image matching and subtraction techniques as implemented in the
ISIS~2.2 image-subtraction package \citep{ala00}.  The 8.0~$\mu$m
subtracted image is shown in Fig. \ref{fig1}(c).  In \citet{mei06} we
demonstrate the applicability of the image-subtraction technique for
{\it Spitzer/IRAC} SN data and investigate its uncertainties.
Aperture photometry of the subtracted images was then carried out
using the Starlink package {\sc gaia} \citep{dra02}.  A circular
aperture of radius $2\farcs25$ was used for the photometry. This
aperture was chosen as a compromise between maximizing the sampled
fraction of source flux (the radius of the first diffraction minimum
at the extreme red end of the 8.0~$\mu$m channel is $\sim2\farcs6$)
and minimizing any extended residual emission in the subtracted image.
Aperture corrections were derived from the {\it IRAC} PSF images
available on the {\it Spitzer} website. The correction factors were
1.23, 1.26, 1.50, and 1.65 for 3.6, 4.5, 5.8, and 8.0~$\mu$m,
respectively. \\

For each measurement, the aperture was centered on the SN image using
a combination of centroid estimates and visual inspection.  The
residual background level was measured using a clipped mean sky
estimator and a concentric sky annulus having respective inner and
outer radii of 1.5 and 2.8 times the aperture radius.  The results are
shown in Table \ref{tab1}.  The uncertainty was determined from the
sky variance within the sky annulus. These error estimates were
confirmed by measuring the variance in the (day~496~--~day~499)
subtracted frame for each band, assuming a similar underlying error in
the unsubtracted day~670 frame, and appropriately combining the two
errors.  These uncertainties are quoted in Table \ref{tab1}. However,
it is likely that additional systematic errors were present due to
image-matching uncertainties.\\

At 3.6~$\mu$m, 4.5~$\mu$m, and 5.8~$\mu$m the flux differences between
the two methods all have less than $3\sigma$ significance (see Table
\ref{tab1}). At 8.0~$\mu$m the difference is over $4\sigma$. As
discussed below, we attribute this significant difference to the
presence of a residual source in the day~670 image.  In Table
\ref{tab1} we also show the PSF-derived {\it IRAC} fluxes obtained by
S06 for day~496/9.  There is reasonable consistency with our PSF
results, although at 8.0~$\mu$m we see a higher flux at a significance
of just under $3\sigma$.  Given the complexity of the field this
difference is, perhaps, not too surprising.

\subsection{Days~670, 678/81 Results}
On day~670, there was no detectable source at or near the SN position
at 3.6~$\mu$m, 4.5~$\mu$m, or 5.8~$\mu$m.  However, sources were
detected near the SN position at 8.0~$\mu$m on day~670 and 24~$\mu$m
on day~678/81 (see Figure \ref{fig1}).  We compared the positions of
these sources with that of the SN.  The coordinates of the SN were
measured by applying PSF fitting (using {\it SNOOPY}) to the
subtracted 8~$\mu$m image (which was in the coordinate system of the
day~496/9 SINGS image).  For this, we used the PSF obtained from the
day 496/9 8~$\mu$m (SINGS) image which has a PSF identical to that of
the subtracted image.  We also measured the SN coordinates in the same
image using three other methods: centroiding, optimal filtering, and
Gaussian fitting as implemented in the IRAF CENTER task. The mean and
standard deviation of the results from these four methods were adopted
as the SN position and uncertainty, respectively. \\

To convert the SN coordinates to the day~670 {\it IRAC} and day~678/81
{\it MIPS} images, we derived geometric transformations between these
images and the day~496/9 8~$\mu$m SINGS image.  The transformation
between the day~496/9 and day~670 8~$\mu$m images was obtained using
the centroid coordinates of 20 isolated sources visible in both
frames. We used IRAF GEOMAP to derive a general transformation
including shifts, scales, and rotations in $x$ and $y$, and a
second-order polynomial for the nonlinear part. The transformation
between the day~496/9 8~$\mu$m SINGS image and the 24~$\mu$m {\it
MIPS} image was obtained in a similar manner, using the centroid
coordinates of 20 isolated sources visible in both frames.  \\

To measure the coordinates of the 8~$\mu$m and 24~$\mu$m sources
detected near the SN position, we again used PSF fitting ({\it
SNOOPY}). The associated uncertainties were estimated by simulating
point sources using a Gaussian PSF with a flux similar to the faint
source. Artificial sources were placed in each of the images at nine
positions where the background was judged to have a similar level and
complexity to that of the SN location. The coordinates of these
sources were measured with PSF fitting using the same polynomial
orders for modelling the background as for the actual 8~$\mu$m and
24~$\mu$m sources.  Finally, the measured coordinates were compared
with the known positions of the simulated sources and the standard
deviations of their offsets were adopted as the uncertainties in our
coordinate measurements. \\

Our conclusion from the above astrometric measurements is that the
8~$\mu$m source coincides with the SN position to within $1\arcsec$
(90\% confidence), and the 24~$\mu$m source coincides with the SN
position to within $2\arcsec$ (90\% confidence) i.e. in both cases the
coincidence is to within 1 native pixel ($1\farcs2$ at 8~$\mu$m,
$2\farcs5$ at 24~$\mu$m). The bulk of the position coincidence
uncertainty arose from the PSF fitting to the days~670-81 sources with
a smaller contribution from the co-ordinate transformation and a
negligible contribution from measuring the position of the SN on
day~496/9. \\

We measured the fluxes of the days~670--81 sources using our
PSF-fitting procedure, obtaining $73 \pm 7~\mu$Jy at 8.0~$\mu$m and
$380 \pm 90~\mu$Jy at 24.0~$\mu$m.  The 8.0~$\mu$m flux is consistent
with the $77 \pm 17~\mu$Jy difference in the flux between the
day~496/9 image and the subtracted image (see Table 1), indicating
that the difference was due to the residual source in the day~670
image.  The sensitivity on day~670 and day~678/81 is dominated by the
effects of the bright nearby sources on the PSF fitting.  For the
other three {\it IRAC} channels, upper limits were obtained based on
direct PSF measurements of the day~670 images and on the difference
between the day~496/9 images and subtracted images.  Our $2\sigma$
upper limits at 3.6~$\mu$m, 4.5~$\mu$m, and 5.8~$\mu$m are,
respectively, $10~\mu$Jy, $20~\mu$Jy, and $35~\mu$Jy, rounded to the
nearest 5~$\mu$Jy.

\subsection{Day~1264 result}
As indicated in Section~2, we were recently given access to the PBCD
24~$\mu$m {\it MIPS} image of SN~2003gd obtained within the Sugerman
et al. {\it Spitzer} program (PID:~30494) on day~1264.  Visual
inspection suggests that the source near the SN position had faded
since day~678/81. To investigate this more quantitatively, we
subtracted the day~1264 image from the day~678/81 data using the
procedures described in \S 3.1.2.

We performed the subtraction on both the {\it SINGS}-processed
day~678/81 image (pixel size $1\farcs5$) and on the two original PBCD
images with the native pixel scale of $2\farcs5$ (an average of the
native scale subtracted images was formed).  A discrete source close
to the SN position was observed in both subtractions.  Aperture
photometry of the source was carried out using a $6\farcs1$ radius
aperture. The background was determined by using concentric sky annuli
in the ratio 1.5:2 of the aperture radius, and also by placing the
aperture (without sky annuli) at a number of positions in a
$2' \times 0.5'$ box centered on the SN. The uncertainty was
estimated from the rms value of the aperture values in the second
method. The whole procedure was then repeated with a $4\farcs4$
aperture radius.  Generally consistent results were obtained. The mean
flux measured was $295\pm70~\mu$Jy.  We conclude that it seems likely
that the day~678/81 24~$\mu$m source faded significantly by day 1264.\\

\section{Analysis}
The mid-IR fluxes for SN~2003gd at day~496/9 (Table \ref{tab1}) are
plotted in Figure \ref{fig2}.  The crossbars give the {\it IRAC}
filter bands and the flux error bars are $1\sigma$.  We show both the
PSF-derived points and those derived by image subtraction.  Also shown
are $BVRI$ points obtained on day~493 \citep{hen05} adjusted to
day~496/9 using the SN~1987A light curves.  All the SN~2003gd points
were dereddened using the \citet{car89} extinction law with $R_V =
3.1$ and $E(B-V) = 0.135$~mag \citep{hen05}. \\

There is clearly a strong mid-IR excess.  The IR excess might be
produced by an IR~echo from circumstellar dust, but S06 argue that the
decline rate is too high to be a typical IR echo. We find that it is,
in fact, possible to reproduce the decline rate using a simple IR~echo
model \citep{mei06} with a modest dust shell, although the shell
parameters have to lie within quite a narrow range.  Without more
extensive temporal coverage, it is not possible to conclusively
eliminate a significant IR~echo contribution to the mid-IR emission at
day~496/9.  However, the observed optical attenuation effects
\citep{hen05,sug06} show that some dust condensation in the ejecta
must have taken place. In addition, there is no sign of radio
emission, implying a paucity of circumstellar matter \citep{vdy03}.
Consideration of the deposited radioactive energy also tends to
support dust condensation at this epoch (see below).  Given these
facts, plus the need for a rather specific CSM shell geometry for an
IR echo to reproduce the decline rate between day~496/9 and day~670,
we proceed on the assumption that the day~496/9 mid-IR flux was
probably dominated by emission from newly formed dust in the ejecta.

\subsection{Comparison of SN~2003gd on Day~496/9 with SN~1987A}
In order to interpret further the day~496/9 mid-IR emission from
SN~2003gd, ideally we would compare its spectral energy distribution
(SED) with similar-epoch spectra from a sample of SNe~II-P, but such a
database covering the 3--9~$\mu$m range does not yet exist. The only
pre-{\it Spitzer} 3--9~$\mu$m CCSN spectra are for SN~1987A.  While
SN~1987A was initially atypical (it arose from a blue supergiant
star), its nebular optical/near-IR behavior has been shown
\citep{poz06} to be more similar to that of a normal SN~II-P such as
SN~2002hh.  In addition, quite similar mid-IR spectral behavior has
been found for SN~1987A and the Type~II-P SNe 2004dj and 2005af around
days~200--250 \citep{kot06}.  Only in the [Ar~III] 6.99~$\mu$m line is
significantly different behavior detected.  We conclude that the
nebular mid-IR behavior of SN~1987A is similar to that of Type~II-P
events like SN~2003gd. \\

For comparison with SN~2003gd, we used SN~1987A spectra at
0.3--1.1~$\mu$m [SUSPECT database and \citet{pun95}], 1.05--4.1~$\mu$m
\citep{mei93}, and 4.3--13.0~$\mu$m \citep{bou93,roc93,woo93}.  We
used data from SN~1987A epochs as follows: optical/day~498,
near-IR/day~494, mid-IR/day~494 \citep{mei93}, day~517 \citep{roc93},
and day~518 \citep{bou93}.  This still left the blue half of the
8.0~$\mu$m band unrepresented. To fill in this gap, which includes the
strong [Ni~II]~6.63~$\mu$m line, we used the day~415 SN~1987A KAO
spectrum \citep{woo93}, scaled and shifted to match the day~517/8
spectra in the overlap regions (4.5--5.3, 7.8--12~$\mu$m).  All the
spectra were dereddened. In addition, to convert the spectra to the
SN~2003gd epochs, small scaling adjustments were made using the
SN~1987A light curves.  The SN~1987A spectra were then scaled by $2.9
\times 10^{-5}$ and 0.21 to compensate, respectively, for the
distance and $^{56}$Ni mass differences between the two SNe. \\

\begin{figure*}
\vspace{-1.5cm}
\epsscale{0.64}
\plotone{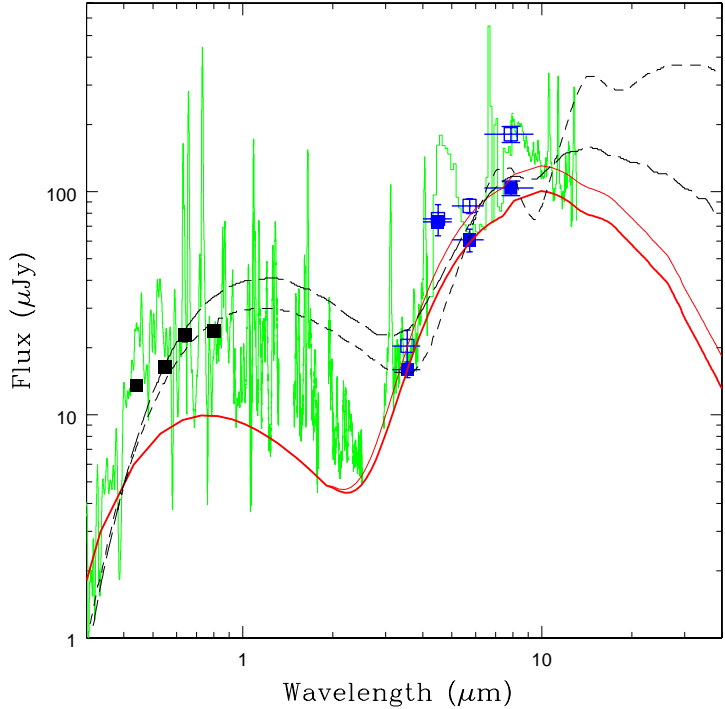}
\vspace{-0.5cm}
\caption[]{ Spectral energy distribution of SN~2003gd on day~496/9
compared with the coeval SN~1987A spectrum and dust emission models.
All data have been dereddened. The spectrum has been corrected for
redshift.  See text for explanation of interpolation of the data to
days~496/9. Open (blue) squares in IR region: PSF-derived {\it IRAC}
points. Solid (blue) squares: image-subtraction-derived {\it IRAC}
points. The cross bars indicate the {\it IRAC} bandwidths. Solid
(black) squares in optical region: photometry from \citet{hen05}.
Faint structured (green) spectrum: optical/IR spectrum of SN~1987A
scaled to allow for differences in distance and $^{56}$Ni mass. The
3--13~$\mu$m spectrum flux has been multiplied by an additional factor
of 2.8. Light continuous line (red): dust emission model matched to
PSF-derived fluxes at 3.6 and 5.8~$\mu$m (this work).  Heavy
continuous line (red): match to image-subtraction-derived fluxes at
3.6 and 5.8~$\mu$m (this work).  Note that the hot component was
adjusted to match the underlying continuum of the scaled SN~1987A
optical spectrum and {\it not} the broad-band points of SN~2003gd,
which would contain a significant contribution from the many strong
emission lines.  Dashed lines (black): dust emission models from
S06. Long-dash: smooth model.  Short-dash: clumped model.
\label{fig2}
}
\end{figure*}

Following all these adjustments, we found that the optical spectrum
showed good consistency with the SN~2003gd photometry at both
epochs. However, to match the SN~2003gd fluxes in the region of the IR
excess (3--13~$\mu$m), we had to further increase the SN~1987A
spectral fluxes by a factor of 2.0 for the image-subtracted points,
and by a factor of 3.1 for the PSF-derived points.  A compromise
factor of 2.8 was applied (Fig. \ref{fig2}). \\

Comparison of the coeval SN~1987A IR spectrum with the day~496/9
SN~2003gd photometry (Fig. \ref{fig2}) shows that much of the {\it
IRAC} fluxes are likely to be due to emission from CO
($\sim$4.8~$\mu$m), SiO ($\sim$8.2~$\mu$m), fine-structure lines, and
Br$~\alpha$. In the Kuiper Airborne Observatory (KAO) study of
SN~1987A \citep{woo93}, barely 20\% of the fluxes corresponding to the
{\it IRAC} 4.5~$\mu$m and 8.0~$\mu$m bands were ascribed to emission
from ejecta dust. On the other hand, in SN~2003gd the 4.5~$\mu$m point
is not as far above the continuum as one might expect given the level
of the SN~1987A CO emission.  A similar, but less pronounced, effect
may be apparent at 8~$\mu$m.  This suggests that, while the factor of
2.8 scaling is appropriate to match the IR continua of SN~1987A to
that of SN~2003gd, it exaggerates the line and molecular emission from
the latter SN. Nevertheless, the non-dust contributions to the
4.5~$\mu$m and 8.0~$\mu$m fluxes of SN~2003gd are probably still
significant, so these points should not be used for matching any dust
emission model.  In contrast, the 3.6~$\mu$m and 5.8~$\mu$m fluxes lie
quite close to the SN~1987A continuum.  Given that this continuum was
due to emission from ejecta dust, we conclude that these points
provide a fair measure of the emission from newly formed dust in
SN~2003gd. We make use of these two points in matching the dust
emission model.

\subsection{Dust Mass at Day 496/9} 
To estimate the dust mass produced in SN~2003gd, we compared a simple
analytical IR-emission model with the observed SEDs.  An additional
component was added to represent continuum emission from hot,
optically thick gas in the ejecta.  To select the likely grain density
distribution and grain materials for the dust emission model, we
sought guidance from dust condensation calculations and the explosion
models upon which they are based.  Only a few papers have been
published which describe SN dust condensation based on explosion
models.  Such papers fall into two categories: SN~1987A, and
high-redshift low-metallicity progenitor SNe. No calculations for
local Type~II-P events have been published. We judge the SN~1987A dust
models as probably being the more relevant. \\

\citet{koz89} and \citet{tod01} have calculated dust condensation
within the ejecta of SN~1987A. These authors used the ejecta chemical
composition as determined in nucleosynthesis models
\citep{has89,nom91}.  Both sets of authors adopted complete chemical
mixing within the dust-forming zone.  Within this zone Todini \&
Ferrara assumed a uniform density distribution.  \citet{koz89} used
the density profile from an explosion model \citep{has89} but this
also is roughly flat.  Similar dust-type abundances were obtained by
both sets of authors, but neither make explicit predictions about the
dust distribution within the ejecta. Recent three-dimensional CCSN 
explosion models \citep{kif06} confirm that extensive mixing of the core 
takes place. The same models also show that the density structure is likely
to be exceedingly complex, with high-density clumps moving out through
lower-density gas.  How this affects the dust distribution has yet to
be determined.  \\

Given the current state of knowledge, we assume that dust of uniform
number density forms throughout the zone containing abundant
refractory elements. The extent of this zone can be assessed using the
late-time widths of metal lines.  In the day~493 optical spectrum of
SN~2003gd \citep{hen05}, the maximum velocities implied by the metal
lines generally do not exceed $\sim$2000~km~s$^{-1}$. This upper limit
is adopted as the size of the dust-forming region.  The uniform
density assumption is conservative in that it provides the least
effective way of hiding dust grains in optically thick regions.
Guided by the dust-formation calculations \citep{koz89,tod01,noz03},
we included silicate, amorphous carbon, and magnetite dust in the mass
ratios 0.68/0.16/0.16. The mass absorption functions for the three
materials were taken from the literature \citep{lao93,rou91,koi81}. \\

Our dust IR-emission model comprises a uniform sphere of isothermal
dust grains.  Following the escape probability formalism
\citep{ost89,luc89}, the luminosity ($L_{\nu}$) of the sphere at 
frequency $\nu$ is given by
\begin{equation}
L_{\nu}= 2\pi^2R^2B_{\nu}(T)[\tau_{\nu}^{-2}(2\tau_{\nu}^2-1+(2\tau_{\nu}+1)e^{-2\tau_{\nu}})],
\end{equation}
where $R$ is the radius of the dust sphere at some time after the
explosion, $B_{\nu}(T)$ is the Planck function at temperature $T$, and
$\tau_\nu$ is the optical depth to the center at frequency $\nu$.  For
a grain size distribution $dn =ka^{-m}da$, where $dn$ is the number
density of grains having radius $a \to a+da$, $m$ is typically between
2 and 4, and $k$ is the grain number density scaling factor, it can
be shown that $\tau_\nu=\frac{4}{3}\pi
k\rho\kappa_{\nu}R\frac{1}{4-m}[a^{4-m}_{max}-a^{4-m}_{min}]$, where $\rho$
and $\kappa_{\nu}$ are, respectively, the density and mass absorption
coefficient of the grain material.  The grain size distribution law
was set at $m=3.5$ \citep{mat77} with $a_{min}=0.005~\mu$m and
$a_{max}=0.05~\mu$m.  The total mass of dust, $M_d$, was then found
from $M_d=4\pi R^2\tau_\nu/3\kappa_\nu$ \citep{luc89}.  \\

The model-free parameters are the grain temperature, sphere radius,
and grain number density scaling factor, $k$.  These were adjusted to
reproduce just the 3.6~$\mu$m and 5.8~$\mu$m points.  \citet{woo93}
showed that during the second year of SN~1987A, the dust-emission
continuum could be contaminated by blackbody emission from hot,
optically thick gas, as well as by free-bound radiation. Here we
represent both effects using a single hot blackbody.  The hot
component was adjusted to match the underlying continuum of the scaled
SN~1987A optical spectrum and {\it not} the broad-band points of
SN~2003gd, which would contain a significant contribution from the many
strong emission lines.  We found that the effect of the hot component
on emission longward of 3~$\mu$m was small.  \\

Model matches to both the PSF-fitting and image-subtraction-derived
fluxes were obtained.  The dust emission models are shown in Figure
\ref{fig2}.  We found that to achieve reasonable matches to the data
it was necessary to increase the dust mass until it was optically
thick in the mid-IR. Consequently, we were unable to derive a unique
dust mass since, as we increase the optical depth, ever larger amounts
of dust can be ``hidden'' with little effect on the observed
radiation. We therefore, conservatively, sought the {\it minimum} dust
mass which would provide a satisfactory match to the data.  The model
parameters including the derived dust masses are given in Table
\ref{tab2}.\\

Dust masses of $6\times10^{-5}$~M$_{\odot}$ (PSF fitting, day~496/9
image) and $4\times10^{-5}$~M$_{\odot}$ (aperture photometry,
subtracted image) were obtained. A 1~Mpc reduction in distance reduces
the masses by about 10\%.  The uniform dust distribution of our model,
optically thick at 10~$\mu$m, would surely extinguish all metal lines
in the optical region. Yet, as late as day~493 \citep{hen05} and
day~521 \citep{sug06}, such lines could still be seen. This implies
that the dust distribution must have been ``clumpy,'' allowing some of
the optical line radiation to escape from the nebula.  The presence of
clumping is confirmed by consideration of the ``covering factor,''
$f$.  This is obtained by dividing the projected area of the model
dust sphere by the projected area corresponding to the estimated
extent of the dust-forming zone (2000~km~s$^{-1}$).  A covering
factor of $\sim$0.15 was obtained (Table \ref{tab2}). This may also
account for the relatively modest extinctions in the $R$ band
\citep{sug06}.  We note that SN~1987A showed strong evidence for dust
clumping \citep{luc91}.  In Section~5, we suggest that, in general, SN
ejecta dust becomes optically thick in the mid-IR when the dust mass
exceeds only a few times $10^{-3}$~M$_\odot$. \\

It is argued above that the 8~$\mu$m point should not be used for
constraining the dust model due to possible contamination by SiO
emission. Nevertheless, we investigated the effect of including this
point and found that satisfactory matches can be obtained using
somewhat lower temperatures and higher radii for the model.  Similar
dust masses are derived. However, the match to the SN~1987A spectrum
redward of 8~$\mu$m is very poor, with the model flux exceeding the
continuum flux by about a factor of two. Given the argument in
\S~4.1 that the nebular behavior of SNe~1987A and 2003gd is
similar, we conclude that SiO is indeed contaminating the SN~2003gd
spectrum.\\

The total luminosity of our dust model (for the match to the
PSF-derived fluxes) plus the estimated total optical/near-IR
contribution (i.e. line/molecular emission plus underlying continuum)
is $2.1 \times 10^{39}$~erg~s$^{-1}$, with roughly 30\% of the
luminosity arising from the dust.  The $^{56}$Ni mass inferred by
\citet{hen05} is 0.016$^{+0.010}_{-0.006}$~M$_{\odot}$.  Dividing the
observed total luminosity by the radioactive decay energy deposited in
the ejecta \citep{li93}, scaled to the $^{56}$Ni mass of SN~2003gd, we
obtain 1.2$^{+0.7}_{-0.5}$. Thus, the total luminosity is similar to
that resulting from the deposited radioactive energy. This tends to
support the proposition that newly condensed ejecta dust was
responsible for the mid-IR continuum emission.  Use of a lower
distance would reduce the radioactive decay energy required to produce
the observed flux. However, this would not significantly affect the
energy constraints as the inferred $^{56}$Ni mass would also fall ---
that is, the {\it fraction} of radioactive decay luminosity required
to produce the observed flux would stay about the same.  The dust
masses we derive are about 25\% of the $2.0\times10^{-4}$~M$_{\odot}$
which S06 obtain from their smooth model fit at the same epoch
(day~496/9).  The S06 models are shown in Fig. \ref{fig2} (dashed
lines).  Between 3~$\mu$m and 10~$\mu$m their models are in
approximate agreement with ours, but at longer wavelengths our model
shows a much sharper decline.  It appears that the S06 model predicts
a component of colder dust and this would account for their larger
dust masses.  Their model invokes a source luminosity of $2.6 \times
10^{39}$~erg~s$^{-1}$. This is $\sim$20\% larger than the value in our
model but is still consistent with the radioactive energy deposited,
given the uncertainties in the $^{56}$Ni mass.

\subsection{The Days~670-681 Sources}
S06 found a 24~$\mu$m flux on day~678/81 of $106 \pm 16~\mu$Jy.  They
also reported upper limits at 3.6~$\mu$m, 4.5~$\mu$m, 5.8~$\mu$m, and
8.0~$\mu$m on day~670.  {\it It is from these later-epoch measurements
that they deduce a dust mass of $0.02$~M$_{\odot}$.}  As indicated
above, we also obtained no detection at 3.6~$\mu$m, 4.5~$\mu$m, and
5.8~$\mu$m on day~670. However, at 8.0~$\mu$m we obtained a
significant detection in our PSF fitting of $73 \pm 7~\mu$Jy.
Moreover, inspection of Fig. \ref{fig1}(b) does appear to confirm the
presence of a source close to the SN position.  At 24~$\mu$m our
measured flux of $380 \pm 90~\mu$Jy (see above) is about a factor of 4
larger than that obtained by S06.  \\

To investigate this flux difference we assessed the day~678/81 {\it
MIPS} sensitivity at the source position using a number of methods.
The complex field in the SN vicinity makes direct noise estimation
quite difficult.  Therefore, to determine the underlying
pixel-to-pixel noise, we subtracted the day~681 image from the day~678
image using the procedures described above.  We then measured the
noise at the SN location.  We used a $6\farcs1$ radius aperture which
encompasses about 0.93 of the flux in the Airy disk at 24~$\mu$m. The
flux in the aperture was measured at a series of locations within
$40\arcsec$ of the SN position.  The rms value, after aperture
correction, is $\sim$200~$\mu$Jy. However, as the subtracted image
contained the noise of the two original images, we divided this by
$\sqrt2$, yielding 140~$\mu$Jy as the intrinsic sensitivity
($1\sigma$) of the {\it MIPS} data.  A further $\sqrt2$ improvement of
the sensitivity to 100~$\mu$Jy arises from the fact that the
SINGS-processed image, used by S06 and ourselves for the PSF fitting,
is a combination of the two SINGS PBCD images.  \\

As a further check, we examined the {\it MIPS} sensitivity in the
SINGS-processed image well away from the galaxy in a relatively
``clean'' part of the sky, lying about $4\arcmin$ south of the SN
location. Artificial stars were placed at 11 different positions
within a $1\arcmin \times 3\arcmin$ area.  The input star flux was set
at 130~$\mu$Jy and the flux at each of the 11 positions was measured
by aperture photometry, using a $6\farcs1$ radius aperture and a sky
annulus between 1.5 and 2 times this radius.  The effective
sensitivity was assessed from the dispersion in the flux values. The
procedure was repeated with an input star flux of 5300~$\mu$Jy.  The
dispersion in both the low-flux and high-flux cases gave about the
same result, indicating that even well away from the galaxy, the
sensitivity was background limited. The $1\sigma$ sensitivity, after
aperture correction, was found to be about 90~$\mu$Jy, similar to the
value obtained from image subtraction.  \\

As a final check on the above procedures, we used the {\it Spitzer
Sensitivity-Performance Estimation Tool (PET)} to estimate the
intrinsic sensitivity. The measured background near the SN is about
equivalent to the ``high background'' setting of the PET. From this we
derive an intrinsic $1\sigma$ sensitivity of 60~$\mu$Jy, of
similar magnitude to the directly-determined sensitivity values.  \\

We conclude that the actual $1\sigma$ sensitivity of the day~678/81
{\it MIPS} image at the SN position was $\sim$90~$\mu$Jy, much larger
than the 16~$\mu$Jy claimed by S06. Moreover, the 106~$\mu$Jy
flux at the SN position claimed by S06 would only yield a S/N
of about unity, not $\sim6.5$ as they reported.
However, we note that scaling the S06 result by a factor of 4 yields
$424 \pm 64$~$\mu$Jy.  This is more consistent with both our flux
value and with our separately measured {\it MIPS} sensitivity.  We
suspect, therefore, that there is an error in the 24~$\mu$m flux
reported by S06. \\

What is the origin of the 8~$\mu$m and 24~$\mu$m sources on
days~670--81?  Given the very crowded field within which the supernova
occurred, and the fact that CCSNe tend to occur near star-forming
regions, a cool background source lying close to the SN might be
considered.  A similar situation was described recently for {\it
Spitzer} observations of the CCSN SN~2002hh \citep{mei06}.  However,
the fading of the 24~$\mu$m source (\S~3.3) tends to rule out a
background source, at least for most of the 24~$\mu$m flux.  We shall
therefore consider the implications of assuming that the sources are
ultimately due to the SN. \\

We first hypothesize that the days~670-81 8~$\mu$m and (unsubtracted)
24~$\mu$m sources have the same origin and that this origin is the SN
ejecta.  A simple blackbody match to our days~670-81 flux measurements
yields a temperature of 250~K, a radius of $1.9\times10^{16}$~cm, and
a luminosity of $9.7\times10^{38}$~erg~s$^{-1}$. This is immediately
problematic.  To attain a radius of $1.9\times10^{16}$~cm the material
at the outer limit of the blackbody would have to be travelling at
3200~km~s$^{-1}$. This is substantially larger than the 2000~km~s$^{-1}$ 
limit on metal velocities indicated by late-time spectra. In addition, 
after adding an additional $1.1\times10^{38}$~erg~s$^{-1}$ due to the
optical/near-IR emission estimated from the optical photometry, we
obtain a total luminosity of $10.8\times10^{38}$~erg~s$^{-1}$.  This
is a factor of 4 more than the total radioactive decay energy
deposited in the ejecta, according to the formula of \citet{li93},
scaled to the $^{56}$Ni mass of SN~2003gd. Indeed, it exceeds the {\it
total} radioactive luminosity (i.e., including escaping gamma rays) by
more than a factor of 2. Even allowing for the uncertainty in the
$^{56}$Ni mass, the energy deficit is severe.  (We note that, even
with their apparently underestimated 24~$\mu$m flux, the day~678/81
luminosity invoked by S06 exceeds the deposited radioactive energy by
$\sim$50\%.)  It is possible for the bolometric luminosity to exceed
that of the instantaneous radioactive decay deposition when the
recombination timescale exceeds the radioactive or expansion
timescales. However, this commences at much later epochs ($>$day~800)
than those considered here \citep{koz98}.  Thus, on both energy and
velocity considerations, we have evidence that most of the 8~$\mu$m
and 24~$\mu$m fluxes cannot be due to emission from supernova ejecta
dust. \\

Let us now suppose that only the fading component of the 24~$\mu$m
source is due to ejecta dust while the remainder of the 24~$\mu$m flux
plus some or all of the 8~$\mu$m flux is due to a background source.
At 250~K, to match the fading component would require a blackbody
luminosity of 3 times the likely deposited radioactive luminosity and
a velocity of $2800^{+300}_{-400}$~km~s$^{-1}$, where the error is due
to the flux uncertainty. Even if we reduce the distance by 1~Mpc and
use the lower limit of the flux values, the velocity still exceeds
2000~km~s$^{-1}$ and the luminosity still exceeds the deposited
radioactive luminosity by a factor of 2.  Reducing the temperature
from 250~K to 150~K, the luminosity falls by 30\% but the velocity of
the blackbody surface rises to an increasingly implausible
$6700^{+700}_{-900}$ km~s$^{-1}$. (We note that, in their model, S06
invoke an outer limit for their dust zone of $\sim$8000~km~s$^{-1}$,
which is even more unlikely.)  Increasing the temperature above 250~K
also does not help since the luminosity deficit problem would
worsen. Moreover, this would produce an 8~$\mu$m flux in excess of
that seen near the SN position.  Similar results are obtained if we
employ our dust emission model rather than a blackbody.  We conclude
that most of the fading component of the 24~$\mu$m flux cannot be due
to dust in the SN ejecta. \\

If the mid-IR flux observed near the position of SN~2003gd is not due
to condensing dust in the ejecta, then what could be the origin of the
emission?  The substantial fading at 24~$\mu$m between day~678/81 and
day~1264 points to a causal connection with the SN. A possible
scenario is that the mid-IR emission originated in an IR echo from
circumstellar or interstellar gas.  As an illustration, we have
estimated the parameters of a dust sheet lying in front of the SN
required to reproduce the fading component of the 24~$\mu$m flux.  We
used an IR~echo model similar to that described by \citet{mei06}.  The
input bolometric light curve was based on the information given by
\citet{hen05}, with a single grain radius of 0.07~$\mu$m. Estimates
were repeated using the specific grain emissivities of different dust
species. Preliminary results suggest that the 24~$\mu$m flux can be
reproduced with a dust sheet of H number density $\sim$10~cm$^{-3}$,
gas-to-dust ratio of 100, lying 10--20~pc in front of the SN.  At this
distance the dust temperature is 75--90~K.  The optical depth to
UV-optical photons is $\sim$0.2.  The echo radius would be about
$0\farcs1$ and so such a source would be effectively coincident with
the SN position.  To account for the fading the dust sheet would have
to be of irregular density on scales of a few parsecs (a fraction of
an arcsecond).  \citet{sug05} and \citet{vdy06} found an optical echo
on day~623 lying at $0\farcs3$ from SN~2003gd, with a strong
concentration to the NW.  They showed that this could be explained by a
dust sheet lying about 100~pc in front of the SN.  The one-sided
appearance of the optical echo suggests that such dust sheets can
indeed exhibit large density fluctuations on a scale of only a few
tenths of an arcsecond. We conclude that an IR~echo may well be
responsible for the variable component of the 24~$\mu$m flux from
SN~2003gd.  Further discussion of the IR~echo hypothesis as applied to
SN~2003gd is beyond the scope of this paper. \\

The key point following from the above discussion is that most of the
mid-IR flux at days~670--681 could not have been due to dust in the
supernova ejecta.  In particular, it suggests that the inference by
S06 of a large mass (0.02~M$_{\odot}$) of ejecta dust is
unjustified.\\

\section{Conclusions}
We have examined late-time mid-IR observations of the
Type~II-P SN~2003gd and find the following.

{\it(i)} By day~496/9, at least $4 \times 10^{-5}$
M$_{\odot}$ of dust had formed in the ejecta of SN~2003gd.  The larger
(factor of $\sim$4) mass indicated by the smooth model of S06 appears
to be due to the presence of a larger component of cold dust, but this
has no direct observational support.  After allowing for differences
in $^{56}$Ni production, we find that the optical flux of SN~2003gd is
similar to the coeval value for SN~1987A, while the 3--9~$\mu$m flux
is almost three times stronger.  This may indicate more efficient dust
production in SN~2003gd.  Nevertheless, there is no evidence at this
epoch that the absolute dust production was unusually high.  The dust
masses and temperatures are similar to those inferred for SN~1987A
\citep{woo93}.  There is also evidence that the dust in SN~2003gd
formed in clumps.  Comparison with coeval spectra of SN~1987A shows
that even as late as day~$\sim$500 the extraction of information about
dust formation from broad-band photometry has to be approached with
caution due to the effects of other emission mechanisms.  This
underlines the desirability of acquiring low-resolution spectra for
such studies, since this would allow correction for forbidden lines
and molecular emission.\\

{\it(ii)} Emission from point-like sources close to the SN position
was detected on day~670 at 8~$\mu$m and day~678/81 at 24~$\mu$m.  The
fading of the 24~$\mu$m source (\S~3.3) tends to rule out a background
source as the origin of the mid-IR emission, at least for most of the
24~$\mu$m flux.  However, energy and velocity considerations also rule
out the ejecta of SN~2003gd as the origin of most of the mid-IR
fluxes.  The inference by S06 of 0.02~M$_{\odot}$ of ejecta dust is
based on a 24~$\mu$m flux which we find is only a quarter to a third
of the true value. But even if we adopt their flux, their claim of
such a large mass of dust is unconvincing.  The large dust mass they
find appears to be a consequence of the low characteristic temperature
in their model.  However, in order that sufficient mid-IR radiation
should escape, it seems that the dust formation zone has to be as
large as $\sim$8000~km~s$^{-1}$, in conflict with the observed metal
line velocities. Also, in spite of the low temperature, the input
luminosity of their model required to reproduce the 24~$\mu$m flux
exceeds that of the likely deposited radioactive decay luminosity.
Had the correct (much larger) 24~$\mu$m flux been used, these
difficulties would have been even greater. We conclude that the
0.02~M$_{\odot}$ of ejecta dust deduced by S06 is unsupported by the
data.  These {\it Spitzer} observations provide no basis for the S06
claim that ``the [dust] condensation efficiency implied by SN~2003gd
is close to the value of 0.2 needed for SNe to account for the dust
content of high-redshift galaxies.''  There is, as yet, no direct
evidence that CCSNe are major dust factories. \\

An additional argument against a large detected mass of ejecta dust in
SN~2003gd is pointed out by the referee.  The mass of dust inferred by
S06 for SN~2003gd is a factor of 25 greater than the maximum amount in
SN~1987A determined by \citet{erc07} using a similar model. These
authors suggest that this implies a much higher condensation
efficiency in SN~2003gd.  Yet, the diminution of the optical light
curve of SN~2003gd shown in S06 about 100~days after dust formation is
only about 1.3 mag, very similar to that for SN~1987A at about
the same epoch after dust formation \citep[e.g.,][]{whi89}.  Also the
blueward shifts of the emission lines are not greater in SN~2003gd than in
SN~1987A \citep[e.g.,][]{dan91}.  Given the apparently very different
estimated dust masses between the two SNe, it is difficult to see how
these observed similarities are possible.  One might argue that in
SN~2003gd the dust is more concentrated to the center, but this is
apparently belied by the shift in the H$\alpha$ lines (S06) which
presumably arise farther out in the envelope.  \\

The goal of determining the true dust production in SNe via the
thermal emission from the grains is very challenging. Even at
wavelengths as long as 24~$\mu$m, it is likely that the dust forming
in SN ejecta would become optically thick long before a universally
significant mass of dust was formed.  For example, consider a uniform
distribution of astronomical silicate grains.  For $\lambda>20~\mu$m,
$\kappa_\nu \approx 1000(\lambda(\mu {\rm m})/20)^{-2}$ \citep{lao93},
where $\kappa_\nu$ is the mass absorption coefficient (cm$^2$
g$^{-1}$) at frequency $\nu$.  If we set $\tau_\nu>3$, where
$\tau_\nu$ is the optical depth to the dust sphere center at frequency
$\nu$, and let the radius of the refractory element zone be as large
as $v = 3000$~km~s$^{-1}$, we obtain from our uniform dust model
\begin{equation}
M_d>1.5\times10^{-3}(\lambda(\mu {\rm m})/20)^2(t{\rm (days)}/600)^2~M_{\odot}.
\end{equation}
\noindent
At 24~$\mu$m, and as late as 2~years after the explosion, the lower
limit for the dust mass would still only be
$3\times10^{-3}$~M$_{\odot}$.  Similar lower limits are obtained for
other grain materials such as amorphous carbon. In general,
$\kappa_\nu$ rises toward shorter wavelengths, producing even smaller
lower limits. Dust measurement at still later epochs becomes
increasingly difficult as the grains cool beyond the sensitivity
limits of even {\it Spitzer}.  \\

Of course, if the grains are arranged in optically thick clumps, then
a large mass of dust could be hidden in the clumps. This problem has
been recognized for many years, as in SN~1987A \citep{luc89,woo93} and
SN~1998S \citep{poz04}. \citet{erc07} have recently shown that if
clumps are optically thick in just the optical/near-IR region, but
thin in the mid-IR, then the dust mass may be constrained by the
observed luminosities in the two wavelength regions.  However, once
the dust becomes optically thick in the mid-IR it is possible to
derive lower limits only.  As explained above, this situation sets in
when the dust mass exceeds only a few times $10^{-3}$~M$_{\odot}$,
well below the cosmologically interesting limit of
$\sim0.1$~M$_{\odot}$.  \\

The value of mid-IR studies of CCSNe, such as the {\it Spitzer} work
described here, is that they can test whether dust condensation
is common in typical events (i.e., SNe~II-P).  This is an essential
step if we are to demonstrate that CCSNe are major sources of
universal dust. But if large masses of dust {\it are} formed in SN
ejecta, the direct measurement of the total masses involved is likely
to require observations at much longer wavelengths.  However, even
with such observations there would remain the challenge of eliminating
the effects of IR echoes.

%% If you wish to include an acknowledgments section in your paper,
%% separate it off from the body of the text using the \acknowledgments
%% command.

%% Included in this acknowledgments section are examples of the
%% AASTeX hypertext markup commands. Use \url without the optional [HREF]
%% argument when you want to print the url directly in the text. Otherwise,
%% use either \url or \anchor, with the HREF as the first argument and the
%% text to be printed in the second.

\acknowledgments We thank the referee for valuable comments and
suggestions.  We are grateful to B.~Sugerman and the {\it SEEDS} team
for making their proprietary data available to us.  We thank J. Scalo
and L. Pan for helpful discussions.  We also thank M.~Regan for
providing details about the SINGS-processed images, and B.~Ercolano
and M.~Barlow for discussions about dust/IR~emission models.  S.M. was
supported by funds from the Participating Organisations of EURYI and
the EC Sixth Framework Programme.  C.L.G. was supported in part by
PPARC Grant PPA/G/S /2003/00040.  R.K. was supported in part by EU RTN
Grant HPRN-CT-2002-00303.  M.P. was supported by PPARC Grant
PPA/G/S/2001/00512.  J.C.W. and A.V.F. were supported in part by NSF
grants AST--0406740 and AST--0607485, respectively. This work is based
on observations made with the {\it Spitzer Space Telescope}, which is
operated by the Jet Propulsion Laboratory, California Institute of
Technology, under a contract with NASA. Support for this work was
provided by NASA through an award (3248) issued by JPL/Caltech.

%% To help institutions obtain information on the effectiveness of their
%% telescopes, the AAS Journals has created a group of keywords for telescope
%% facilities. A common set of keywords will make these types of searches
%% significantly easier and more accurate. In addition, they will also be
%% useful in linking papers together which utilize the same telescopes
%% within the framework of the National Virtual Observatory.
%% See the AASTeX Web site at http://www.journals.uchicago.edu/AAS/AASTeX
%% for information on obtaining the facility keywords.

%% After the acknowledgments section, use the following syntax and the
%% \facility{} macro to list the keywords of facilities used in the research
%% for the paper.  Each keyword will be checked against the master list during
%% copy editing.  Individual instruments or configurations can be provided 
%% in parentheses, after the keyword, but they will not be verified.

Facilities: \facility{Spitzer Space Telescope,~} \facility{SSC Leopard
Archive Tool,~} \facility{ NASA/IPAC Extragalactic Database ({\it NED}).}

%% Appendix material should be preceded with a single \appendix command.
%% There should be a \section command for each appendix. Mark appendix
%% subsections with the same markup you use in the main body of the paper.

%% Each Appendix (indicated with \section) will be lettered A, B, C, etc.
%% The equation counter will reset when it encounters the \appendix
%% command and will number appendix equations (A1), (A2), etc.

\appendix

\section{SN Dust Production Required to Account for Observed High-Redshift 
Dust} Models of dust formation in CCSNe \citep{tod01,noz03} succeed in
producing copious amounts of dust --- around 0.1--1~M$_{\odot}$ even
in the low-metallicity environments of the early universe.  This
corresponds to a SN dust condensation efficiency of about 0.2
\citep{mor03}, where the efficiency is defined as the dust mass
divided by the total mass of refractory elements.  This is enough to
account for the quantity of dust seen at high redshifts.  \\

As a demonstration, let us consider the results of \citet{ber03}. In
their study of high-redshift quasars they deduced, from the far-IR
luminosities, a star-formation rate of $\sim$3000~M$_{\odot}$
yr$^{-1}$ and a dust-formation rate of $\sim$1~M$_{\odot}$
yr$^{-1}$. Consider a simple stellar mass spectrum \\
\begin{equation}
dN=\gamma M^{-2.5}dM, \\
\end{equation} 
\noindent
where $dN$ is the number of stars in the mass interval $M$ to $M +
dM$, $\gamma$ is a constant with units of (mass)$^{1.5}$, and all
stars lie within the mass range 0.2~M$_{\odot} < M <$ 30~M$_{\odot}$.
The total stellar mass in the interval $M$ to $M + dM$ is given by
$dM_{tot} = MdN =\gamma M^{-1.5}dM$.  Integrating this equation over
the stellar mass range we obtain $M_{tot}=4.1\gamma$. In one year we
have $M_{tot}=4.1\gamma = 3000$~M$_{\odot}$, so $\gamma=732$.  Stars
of mass exceeding about 8~M$_{\odot}$ will end their lives as
CCSNe. Integrating equation~(A1) over the range 8~M$_{\odot}$ to
30~M$_{\odot}$, we obtain $N = 0.0381\gamma = 28.0$ --- that is, about
28~CCSNe per year would occur. To produce 1~M$_{\odot}$ of dust per
year, the average dust yield of each SN would have to be
0.036~M$_{\odot}$. Thus, the production rates of dust condensation
models \citep{tod01,noz03} are indeed sufficient to account for the
high-redshift dust. \\

The progenitor of SN~2003gd had a mass in the range 6--12~M$_{\odot}$
\citep{vdy03,hen05}. This would produce about 0.3~M$_{\odot}$ of refractory
elements \citep{woo95}. Consequently, for such a SN to match the
required average dust production, the refractory elements would have
to be converted into dust with an efficiency of about 0.1.

%% The reference list follows the main body and any appendices.
%% Use LaTeX's thebibliography environment to mark up your reference list.
%% Note \begin{thebibliography} is followed by an empty set of
%% curly braces.  If you forget this, LaTeX will generate the error
%% "Perhaps a missing \item?".
%%
%% thebibliography produces citations in the text using \bibitem-\cite
%% cross-referencing. Each reference is preceded by a
%% \bibitem command that defines in curly braces the KEY that corresponds
%% to the KEY in the \cite commands (see the first section above).
%% Make sure that you provide a unique KEY for every \bibitem or else the
%% paper will not LaTeX. The square brackets should contain
%% the citation text that LaTeX will insert in
%% place of the \cite commands.

%% We have used macros to produce journal name abbreviations.
%% AASTeX provides a number of these for the more frequently-cited journals.
%% See the Author Guide for a list of them.

%% Note that the style of the \bibitem labels (in []) is slightly
%% different from previous examples.  The natbib system solves a host
%% of citation expression problems, but it is necessary to clearly
%% delimit the year from the author name used in the citation.
%% See the natbib documentation for more details and options.

\clearpage

%% Tables may also be prepared as separate files. See the accompanying
%% sample file table.tex for an example of an external table file.
%% To include an external file in your main document, use the \input
%% command. Uncomment the line below to include table.tex in this
%% sample file. (Note that you will need to comment out the \documentclass,
%% \begin{document}, and \end{document} commands from table.tex if you want
%% to include it in this document.)

%% \input{table}
%\clearpage

\begin{table}
\tablenum{1}
\begin{center}
\caption{ Mid-IR Photometry on Day~496/9 at the Position of SN~2003gd} 
\begin{tabular}{lcccc}
&&&& \\ \hline
 Author/method & \multicolumn{4}{c}{Flux ($\mu$Jy)} \\ 
                    & 3.6~$\mu$m & 4.5~$\mu$m & 5.8~$\mu$m& 8.0~$\mu$m \\ \hline
This work: PSF fit. & 19.9(3.6) & 74(12)    & 85.5(6.5) & 180(15)   \\
This work: Im. sub. & 15.6(1.3) & 72.2(2.3) & 60.2(7.1) & 103.2(7.7)\\
S06: PSF fit.       & 20.8(2.6) & 73.8(5.6) & 64.9(7.3) & 103(22)  \\ 
\hline
\end{tabular}
\tablenotetext{}{ Col.~1 gives the photometry method used to derive
the fluxes. The ``Im. Sub.'' values were obtained by aperture
photometry of subtracted images (see text).  Uncertainties ($1\sigma$) are
shown in brackets.  The fluxes obtained by S06 for this epoch are
shown for comparison.}
\end{center}
\label{tab1}
\end{table}

\clearpage

\begin{table}
\tablenum{2}
\begin{center}
\caption{Model Parameters for Day~496/9.}  
\begin{tabular}{lccccc}
&&&&& \\ \hline
Method & $T_{\rm dust}$ & $\tau_{10~\mu {\rm m}}$& Dust Mass & $f$  & $T_{\rm hot}$\\ 
       & (K)        &                 & M$_{\odot}$       &      & (K)   \\ \hline
PSF fit. & 525      &   2.6           & $6\times10^{-5}$& 0.17 &   6700   \\   
Im sub.  & 525      &   2.3           & $4\times10^{-5}$& 0.13 &   6700   \\  
\hline
\end{tabular}
\tablenotetext{}{Col.~1 gives the photometry method used to derive the
fluxes to which the model was matched.  Optical depths to the center
at 10~$\mu$m are shown in col.~3. The dust masses in col.~4 were
derived assuming a distance of 9.3~Mpc. A 1~Mpc reduction in distance
reduces the masses by about 10\%.  In col.~5, a ``covering factor''
$f$ is shown (see text). Col.~6 gives the adopted temperature of the
hot component. }
\end{center}
\label{tab2}
\end{table}

\clearpage

%% Use the figure environment and \plotone or \plottwo to include
%% figures and captions in your electronic submission.
%% To embed the sample graphics in
%% the file, uncomment the \plotone, \plottwo, and
%% \includegraphics commands
%%
%% If you need a layout that cannot be achieved with \plotone or
%% \plottwo, you can invoke the graphicx package directly with the
%% \includegraphics command or use \plotfiddle. For more information,
%% please see the tutorial on "Using Electronic Art with AASTeX" in the
%% documentation section at the AASTeX Web site,
%% http://www.journals.uchicago.edu/AAS/AASTeX.
%%
%% The examples below also include sample markup for submission of
%% supplemental electronic materials. As always, be sure to check
%% the instructions to authors for the journal you are submitting to
%% for specific submissions guidelines as they vary from
%% journal to journal.

%% This example uses \plotone to include an EPS file scaled to
%% 80% of its natural size with \epsscale. Its caption
%% has been written to indicate that additional figure parts will be
%% available in the electronic journal.

%FIGURE 1

\clearpage

%FIGURE 2

%% If you are not including electronic art with your submission, you may
%% mark up your captions using the \figcaption command. See the
%% User Guide for details.
%%
%% No more than seven \figcaption commands are allowed per page,
%% so if you have more than seven captions, insert a \clearpage
%% after every seventh one.

%% Tables should be submitted one per page, so put a \clearpage before
%% each one.

%% Two options are available to the author for producing tables:  the
%% deluxetable environment provided by the AASTeX package or the LaTeX
%% table environment.  Use of deluxetable is preferred.
%%

%% Three table samples follow, two marked up in the deluxetable environment,
%% one marked up as a LaTeX table.

%% In this first example, note that the \tabletypesize{}
%% command has been used to reduce the font size of the table.
%% We also use the \rotate command to rotate the table to
%% landscape orientation since it is very wide even at the
%% reduced font size.
%%
%% Note also that the \label command needs to be placed
%% inside the \tablecaption.

%% This table also includes a table comment indicating that the full
%% version will be available in machine-readable format in the electronic
%% edition.

%% The following command ends your manuscript. LaTeX will ignore any text
%% that appears after it.


\begin{thebibliography}{}
\bibitem[Alard (2000)]{ala00}Alard, C. 2000, \aap, 144, 363 
\bibitem[Arnett \etal (1989)]{arn89}Arnett, W. D., Bahcall, J. N., 
  Kirshner, R. P., \& Woosley, S. E. 1989, ARAA, 27, 629
\bibitem[Bertoldi \etal (2003)]{ber03} Bertoldi, F., Carilli, C. L.,
Cox, P., Fan, X., Strauss, M. A., Beelen, A., Omont, A., \& Zylka,
R. 2003, \aap, 406, L55
\bibitem[Bouchet \& Danziger (1993)]{bou93}Bouchet, P., \& Danziger,
I. J.  1993, \aap, 273, 45
\bibitem[Bouchet \etal (1991)]{bou91}Bouchet, P., Phillips, M. M.,
Suntzeff, N. B., Gouiffes, C., Hanuschik, R. W., \& Wooden, D. H. 1991,
\aap, 245, 490
\bibitem[Cernuschi, Marsicano, \& Codina (1967)]{cer67}Cernuschi, F.,
Marsicano, F. R., \& Codina, S. 1967, Ann. d'Astr., 30, 1039 
\bibitem[Clayton \etal (1997)]{cla97}Clayton, D. D., Amari, S., \&
Zinner, E. 1997, \apss, 251, 355
\bibitem[Cardelli, Clayton, \& Mathis (1989)]{car89}Cardelli, J. A.,
Clayton, G. C., \& Mathis, J. S. 1989, \apj, 345, 245 
\bibitem[Danziger \etal (1989)]{dan89}Danziger, I. J., Gouiffes, C.,
Bouchet, P., \& Lucy, L. B. 1989, IAU Circ., 4746, 1
\bibitem[Danziger et al. (1991)]{dan91}Danziger, I.J., Lucy, L.B.,
Bouchet, P., Gouiffes, C., in Supernovae, ed. S. E. Woosley
(New York: Springer), 69
\bibitem[Douvion, Lagage, \& Pantin (2001)]{dou01}Douvion, T.,
Lagage, P. O., \& Pantin, E. 2001, \aap, 369, 589 
\bibitem[Draper, Gray, \& Berry (2002)]{dra02}Draper, P. W., Gray, N.,
\& Berry, D. S. 2002, Starlink User Note 214.10
\bibitem[Dunne \etal (2003)]{dun03}Dunne, L., Eales, S., Ivison, R.,
Morgan, H., \& Edmunds, M. 2003, \nat, 424, 285
\bibitem[Dwek(1998)]{dwe98}Dwek, E. 1998, \apj, 501, 643
\bibitem[Dwek \etal (1987)]{dwe87}Dwek, E., Dinerstein, H. L.,
Gillett, F. C., Hauser, M. G., \& Rice, W. L. 1987, \apj, 315, 571 
\bibitem[Dwek \etal (1992)]{dwe92}Dwek, E., Moseley, S. H., Glaccum,
W., Graham, J. R., Loewenstein, R. F., Silverberg, R. F., \& Smith,
R. K. 1992, \apjl, 389, L21
\bibitem[Elmhamdi \etal (2003)]{elm03}Elmhamdi, A., \etal 2003, \mnras,
338, 939
\bibitem[Ercolano, Barlow, \& Sugerman (2007)]{erc07}Ercolano, B.,
Barlow, M. J., \& Sugerman, B. E. K. 2007, \mnras, 375, 753 
\bibitem[Evans \& McNaught (2003)]{eva03}Evans, R., \& McNaught, R.
2003, IAU Circ., 8150, 2
\bibitem[Fall, Charlot, \& Pei (1996)]{fal96}Fall, S. M., Charlot, S.,
\& Pei, Y. C. 1996, \apjl, 464, L43 
\bibitem[Fall, Pei, \& McMahon (1989)]{fal89}Fall, S. M., Pei, Y. C., \&
McMahon, R. G. 1989, \apjl, 341, L5
\bibitem[Garnavich \& Bass (2003)]{gar03}Garnavich, P., \& Bass,
E. 2003, IAU Circ., 8150, 3
\bibitem[Gehrz (1989)]{geh89}Gehrz, R. D. 1989, in Interstellar Dust:
Proceedings of the 135th Symposium of the International Astronomical
Union, ed.  ed. L. J. Allamandola, A. G. G. M. Tielens (Dordrecht:
Kluwer), 445
\bibitem[Gerardy \etal (2002)]{ger02}Gerardy, C. L., Fesen, R. A.,
Nomoto, K., Garnavich, P. M., Jha, S., Challis, P. M., Kirshner, R. P.,
H\"oflich, P., \& Wheeler, J. C. 2002, \apj, 575, 1003 
\bibitem[Hashimoto, Nomoto, \& Shigeyama (1989)]{has89}Hashimoto, M.,
Nomoto, K., \&  Shigeyama, T. 1989, \aap, 210, 5
\bibitem[Hendry \etal(2005)]{hen05}Hendry, M. A., \etal 2005, \mnras,
359, 906
\bibitem[Hoyle \& Wickramasinghe (1970)]{hoy70}Hoyle, F., \&
Wickramasinghe, N. C. 1970, \nat, 226, 62 
\bibitem[Kennicutt \etal (2003)]{ken03}Kennicutt, R. C., Jr., \etal
 2003, \pasp, 115, 928
\bibitem[Kifonidis \etal (2006)]{kif06}Kifonidis, K., Plewa, T.,
Scheck, L., Janka, H.-Th., \& M\"uller, E. 2006, \aap, 453, 661
\bibitem[Kitaura, Janka, \& Hillebrandt (2006)]{kit06}Kitaura, F. S.,
Janka, H.-Th., \& Hillebrandt, W. 2006, \aap, 450, 345
\bibitem[Koike \etal (1981)]{koi81}Koike, C., Hasegawa, H., Asada,
N., \& Hattori, T. 1981, \apss, 79, 77
\bibitem[Kotak \etal (2003)]{kot03}Kotak, R., Meikle, W. P. S.,
Smartt, S. J., \& Benn, C. 2003, IAU Circ., 8152, 1
\bibitem[Kotak \etal (2006)]{kot06}Kotak, R., \etal 2006, \apj, 651,
  117
\bibitem[Kozasa, Hasegawa, \& Nomoto (1989)]{koz89}Kozasa, T., 
Hasegawa, H., \& Nomoto, K. 1989, \apj, 344, 325 
\bibitem[Kozma \& Fransson (1998)]{koz98}Kozma, C., \& Fransson,
C. 1998, \apj, 496, 946
\bibitem[Krause \etal (2004)]{kra04}Krause, O., Birkmann, S. M.,
Rieke, G. H., Lemke, D., Klaas, U., Hines, D. C., \& Gordon, K. D.
2004, \nat, 432, 596
\bibitem[Lagage \etal (1996)]{lag96}Lagage, P. O., Claret, A., Ballet,
J., Boulanger, F., C\'esarsky, C. J., C\'esarsky, D., Fransson, C., \&
Pollock, A.  1996, \aap, 315, L273
\bibitem[Laor \& Draine (1993)]{lao93}Laor, A., \& Draine, B. T. 1993,
\apj, 402, 441
\bibitem[Li, McCray, \& Sunyaev (1993)]{li93}Li, H., McCray, R., \&
Sunyaev, R. A. 1993, \apj, 419, 824 
\bibitem[Lucy \etal (1989)]{luc89}Lucy, L. B., Danziger, I. J.,
Gouiffes, C., \& Bouchet, P.  1989, in Structure and Dynamics of
the Interstellar Medium, ed. G. Tenorio-Tagle, \etal
(Berlin: Springer-Verlag), 164
\bibitem[Lucy \etal(1991)]{luc91}Lucy, L. B., Danziger, I. J., Gouiffes,
C., \& Bouchet, P. 1991, in {\it Supernovae}, ed. S. E.~Woosley (New
York: Springer), 82
\bibitem[Mathis, Rumpl, \& Nordsieck (1977)]{mat77}Mathis, J. S.,
Rumpl, W., \& Nordsieck, K. H. 1977, \apj, 217, 425 
\bibitem[Meikle \etal (1993)]{mei93}Meikle, W. P. S., Spyromilio, J.,
Allen, D. A., Varani, G.-F., \& Cumming, R. J. 1993, \mnras, 261, 535
\bibitem[Meikle \etal (1989)]{mei89}Meikle, W. P. S., Spyromilio, J.,
Varani, G.-F., \& Allen, D. A. 1989, \mnras, 238, 193 
\bibitem[Meikle \etal (2006)]{mei06}Meikle, W. P. S., \etal 2006, 
\apj, 649, 332
\bibitem[Morgan \& Edmunds (2003)]{mor03}Morgan, H. L., \& Edmunds,
M. G. 2003, \mnras, 343, 427
\bibitem[Nomoto \etal (1991)]{nom91}Nomoto, K., Shigeyama, T.,
Kumagai, S., \& Yamaoka, H. 1991, in {\it Supernovae}, ed. S. E. Woosley
(New York: Springer), 176
\bibitem[Nomoto, Sugimoto, \& Sparks (1982)]{nom82}Nomoto, K., Sugimoto, D., \&
Sparks, W. M. 1982, \nat, 299, 803 
\bibitem[Nozawa \etal (2003)]{noz03}Nozawa, T., Kozasa, T., Umeda, H.,
Maeda, K., \& Nomoto, K. 2003, \apj, 598, 78
\bibitem[Osterbrock (1989)]{ost89}Osterbrock, D. E. 1989, 
Astrophysics of Gaseous Nebulae and Active Galactic Nuclei (Mill
Valley, CA: University Science Books)
\bibitem[Pei, Fall, \& Bechtold (1991)]{pei91}Pei, Y. C., Fall, S. M., \&
Bechtold, J. 1991, \apj, 378, 6
\bibitem[Pettini \etal (1997)]{pet97}Pettini, M., King, D. L.,
Smith, L. J., \& Hunstead, R. W. 1997, \apj, 478, 536 
\bibitem[Phillips \etal (2003)]{phi03}Phillips, M., Navarrete, M.,  \&
Preston, G. 2003, IAU Circ., 8152, 2
\bibitem[Pozzo \etal(2004)]{poz04}Pozzo, M., Meikle, W. P. S., Fassia,
A., Geballe, T., Lundqvist, P., Chugai, N. N., \& Sollerman, J.  2004,
\mnras, 352, 457
\bibitem[Pozzo \etal (2006)]{poz06}Pozzo, M., \etal 2006, \mnras, 368,
1169
\bibitem[Pun \etal (1995)]{pun95}Pun, J., \etal 1995, \apjs, 99, 223
\bibitem[Regan \etal (2004)]{reg04}Regan, M. W., \etal 2004, \apjs,
154, 204
\bibitem[Roche, Aitken, \& Smith (1993)]{roc93}Roche, P. F., 
Aitken, D. K., \& Smith, C. H. 1993, \mnras, 261, 522 
\bibitem[Rouleau \& Martin (1991)]{rou91}Rouleau, R., \& Martin,
P. G. 1991, \apj, 377, 526
\bibitem[Smartt \etal (2004)]{sma04}Smartt, S. J., Maund, J. R., Hendry,
M. A., Tout, C. A., Gilmore, G. F., Mattila, S., \& Benn, C. R. 2004,
Science, 303, 499
\bibitem[Sugerman(2005)]{sug05}Sugerman, B. E. K. 2005, \apjl, 632, L17
\bibitem[Sugerman \etal (2006)]{sug06}Sugerman, B. E. K., \etal 2006, 
Science, 313, 196 [S06]
\bibitem[Suntzeff \& Bouchet (1990)]{sun90}Suntzeff, N. B., \&
Bouchet, P. 1990, \aj, 99, 650
\bibitem[Temim \etal (2006)]{tem06}Temim, T., \etal 2006, \aj, 132,
1610 
\bibitem[Tielens (1990)]{tie90}Tielens, A. G. G. M. 1990, NASA
Conf. Pub. 3061, ed. J. C. Tarter, S. Chang, \& D. J. Defrees
(Washington DC: NASA), 59
\bibitem[Todini \& Ferrara(2001)]{tod01} Todini, P., \& Ferrara, A.
2001, \mnras, 325, 726
\bibitem[Van Dyk, Li, \& Filippenko (2003)]{vdy03}Van Dyk, S. D., 
Li, W., \& Filippenko, A. V. 2003, \pasp, 115, 1289
\bibitem[Van Dyk, Li, \& Filippenko (2006)]{vdy06}Van Dyk, S. D., 
Li, W., \& Filippenko, A. V. 2006, \pasp, 118, 351
\bibitem[Whitelock \etal (1988)]{whi88}Whitelock, P. A., \etal 1988,
\mnras, 234, 5P
\bibitem[Whitelock \etal (1989)]{whi89}Whitelock, P. A., \etal 1989,
\mnras, 240, 7
\bibitem[Wooden \etal (1993)]{woo93}Wooden, D. H., Rank, D. M.,
Bregman, J. D., Witteborn, F. C., Tielens, A. G. G. M., Cohen, M.,
Pinto, P. A., \& Axelrod, T. S., 1993, \apjs, 88, 477
\bibitem[Woosley \& Weaver (1995)]{woo95}Woosley, S. E., \& Weaver,
T. A. 1995, \apjs, 101, 181
\end{thebibliography}
\end{document}